\documentclass{nature_mod}

\usepackage{url}
\usepackage{epsfig}
\usepackage{aas_macros}
\usepackage{amssymb,amsmath}

\newcommand{\psr}{J0337$+$1715}
\newcommand{\Sun}{\odot}
\newcommand{\rsun}{R$_\odot$}
\newcommand{\msun}{M$_\odot$}
\newcommand{\degree}{^\circ}

\long\def\symbolfootnote[#1]#2{\begingroup%
\def\thefootnote{\fnsymbol{footnote}}\footnote[#1]{#2}\endgroup} 

\title{A millisecond pulsar in a stellar triple system}

\author{S.~M.~Ransom$^{1}$, I.~H.~Stairs$^{2}$,
  A.~M.~Archibald$^{3,4}$, J.~W.~T.~Hessels$^{3,5}$,
  D.~L.~Kaplan$^{6,7}$, M.~H.~van~Kerkwijk$^{8}$, J.~Boyles$^{9,10}$,
  A.~T.~Deller$^{3}$, S.~Chatterjee$^{11}$, A.~Schechtman-Rook$^{7}$,
  A.~Berndsen$^{2}$, R.~S.~Lynch$^{4}$, D.~R.~Lorimer$^{9}$,
  C.~Karako-Argaman$^{4}$, V.~M.~Kaspi$^{4}$,
  V.~I.~Kondratiev$^{3,12}$, M.~A.~McLaughlin$^{9}$, J.~van
  Leeuwen$^{3,5}$, R.~Rosen$^{1,9}$, M.~S.~E.~Roberts$^{13,14}$,
  K.~Stovall$^{15, 16}$}

\begin{document}

\maketitle

\begin{affiliations}
 \item National Radio Astronomy Observatory, Charlottesville, VA, USA
 \item Dept. of Physics and Astronomy, University of British Columbia, Vancouver, BC, Canada
 \item Netherlands Institute for Radio Astronomy (ASTRON), Dwingeloo,
   The Netherlands
 \item Dept. of Physics, McGill University, Montreal, QC, Canada
 \item Astronomical Institute ``Anton Pannekoek,'' Univ.~of Amsterdam,
   Amsterdam, The Netherlands
 \item Physics Dept., University of Wisconsin-Milwaukee, Milwaukee, WI, USA
 \item Dept.~of Astronomy, University of Wisconsin-Madison, Madison, WI, USA
 \item Dept. of Astronomy and Astrophysics, University of Toronto, Toronto, ON, Canada
 \item Dept. of Physics and Astronomy, West Virginia University, Morgantown, WV, USA
 \item Physics and Astronomy Dept., Western Kentucky University, Bowling Green, KY, USA
 \item Center for Radiophysics and Space Research, Cornell University, Ithaca, NY, USA
 \item Astro Space Center of the Lebedev Physical Institute, Moscow, Russia
 \item Eureka Scientific Inc., Oakland, CA, USA
 \item Physics Dept., New York University at Abu Dhabi, Abu Dhabi, UAE
 \item Physics Dept., University of Texas at Brownsville, Brownsville, TX, USA
 \item Physics and Astronomy Dept., University of New Mexico, Albuquerque, NM, USA
\end{affiliations}

\begin{abstract}
  Gravitationally bound three-body systems have been studied for
  hundreds of years\cite{principia,gut98} and are common in our
  Galaxy\cite{ttsu06,rdl13}.  They show complex orbital interactions,
  which can constrain the compositions, masses, and interior
  structures of the bodies\cite{fab11} and test theories of
  gravity\cite{kv04}, if sufficiently precise measurements are
  available.  A triple system containing a radio pulsar could provide
  such measurements, but the only previously known such system,
  B1620$-$26\cite{tacl99,srh03} (with a millisecond pulsar, a white
  dwarf, and a planetary-mass object in an orbit of several decades),
  shows only weak interactions.  Here we report precision timing and
  multi-wavelength observations of PSR~\psr, a millisecond pulsar in a
  hierarchical triple system with two other stars.  Strong
  gravitational interactions are apparent and provide the masses of
  the pulsar (1.4378(13)\,\msun, where \msun\ is the solar mass and
  the parentheses contain the uncertainty in the final decimal places)
  and the two white dwarf companions (0.19751(15)\,\msun\ and
  0.4101(3)\,\msun), as well as the inclinations of the orbits (both
  $\sim$39.2$\degree$).  The unexpectedly coplanar and nearly circular
  orbits indicate a complex and exotic evolutionary past that differs
  from those of known stellar systems.  The gravitational field of the
  outer white dwarf strongly accelerates the inner binary containing
  the neutron star, and the system will thus provide an ideal
  laboratory in which to test the strong equivalence principle of
  general relativity.
\end{abstract}

Millisecond pulsars (MSPs) are neutron stars that rotate hundreds of
times per second and emit radio waves in a lighthouse-like fashion.
They are thought to form in binary systems\cite{bvdh91} and their
rotation rates and orbital properties can be measured with exquisite
precision via the unambiguous pulse-counting methodology known as
pulsar timing.  As part of a large-scale pulsar
survey\cite{blr13,lbr13} with the Green Bank Telescope (GBT), we have
discovered the only known MSP in a stellar triple system.  The pulsar
has a spin period of 2.73\,ms, is relatively bright ($\sim$2\,mJy at
1.4\,GHz), and has a complex radio pulse profile with multiple narrow
components.

Though initial timing observations showed a seemingly typical binary
MSP system with a 1.6-day circular orbit and a 0.1$-$0.2\,\msun\ white
dwarf (WD) companion, large timing systematics quickly appeared,
strongly suggesting the presence of a third body. There are two other
MSPs known to have multiple companions: the famous pulsar B1257$+$12
which hosts at least 3 low-mass planets\cite{wf92,wol94}, and the MSP
triple system B1620$-$26 in globular cluster M4 with a WD inner
companion and a roughly Jupiter-mass outer
companion\cite{tacl99,srh03}.  The timing perturbations from \psr\
were much too large to be caused by a planetary mass companion.

We began an intensive multi-frequency radio timing campaign (Methods)
using the GBT, the Arecibo telescope, and the Westerbork Synthesis
Radio Telescope (WSRT) to constrain the system's position, orbital
parameters, and the nature of the third body.  At Arecibo, we achieve
median arrival time uncertainties of 0.8\,$\mu$s in 10 seconds,
implying that half-hour integrations provide $\sim$100\,ns precision,
making \psr\ one of the highest-timing-precision MSPs known.

\begin{figure}
  \centerline{\epsfig{angle=0.0,width=6in,file=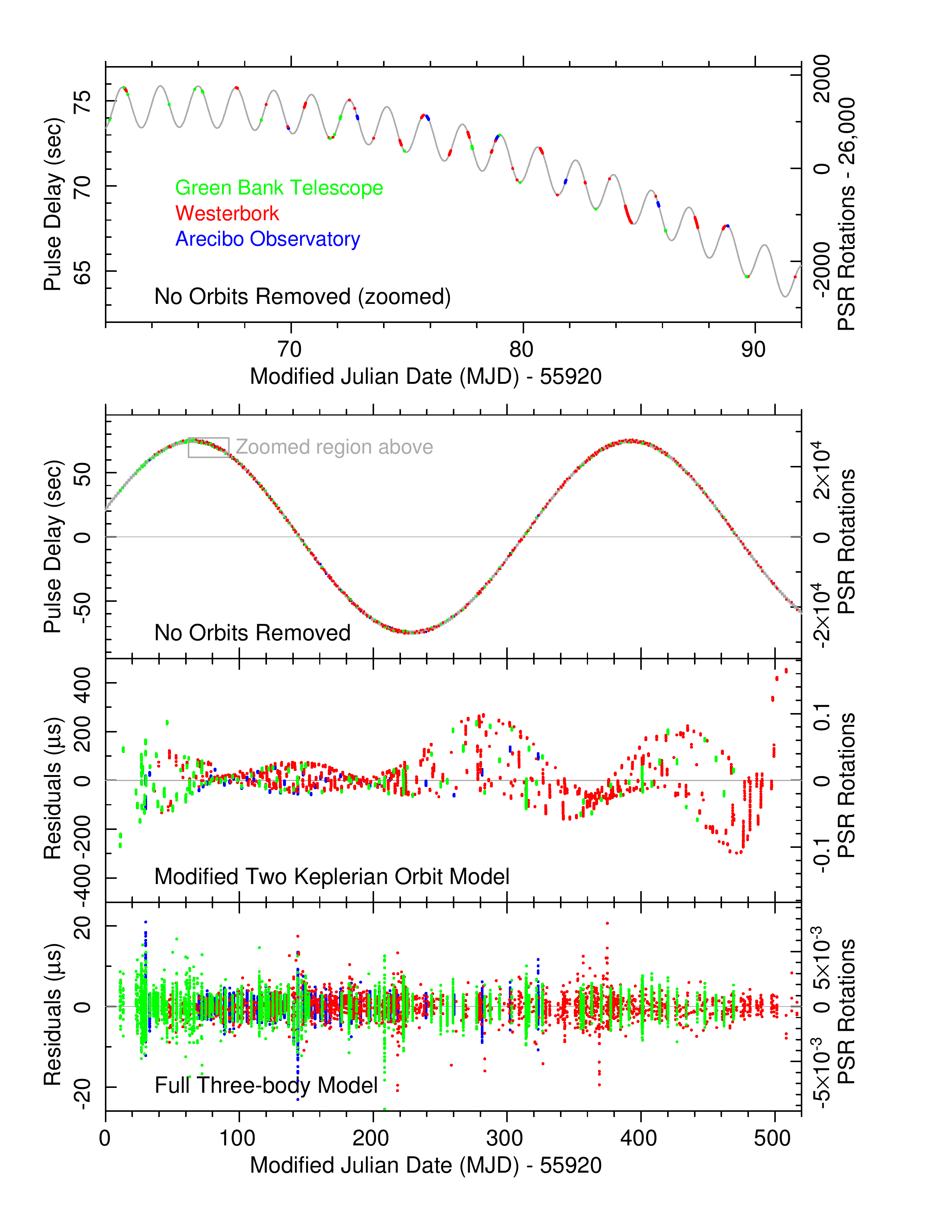}}
  \caption{Timing residuals and delays from the PSR~\psr\ system.  The
    top two panels show geometric light-travel time delays (that is,
    R\o mer delays) in both time and pulse periods, across the inner
    (top panel) and outer (second panel down) orbits, and modified
    Julian dates (MJD) of radio timing observations from the GBT,
    WSRT, and the Arecibo telescope.  Arrival time errors in these
    panels are approximately a million times too small to see.  The
    third panel from the top shows the Newtonian three-body
    perturbations compared with the modified two-Keplerian-orbit model
    used for folding our data at the observed pulse period.  The
    bottom panel shows the post-fit timing residuals from our full
    Markov chain Monte Carlo (MCMC)-derived three-body timing solution
    described in Table\ref{tab:params}.  The weighted root mean
    squared value of the 26,280 residuals is 1.34\,$\mu$s.
    \label{fig:resids}}
\end{figure}

\begin{figure}
  \centerline{\epsfig{angle=0.0,width=\hsize,file=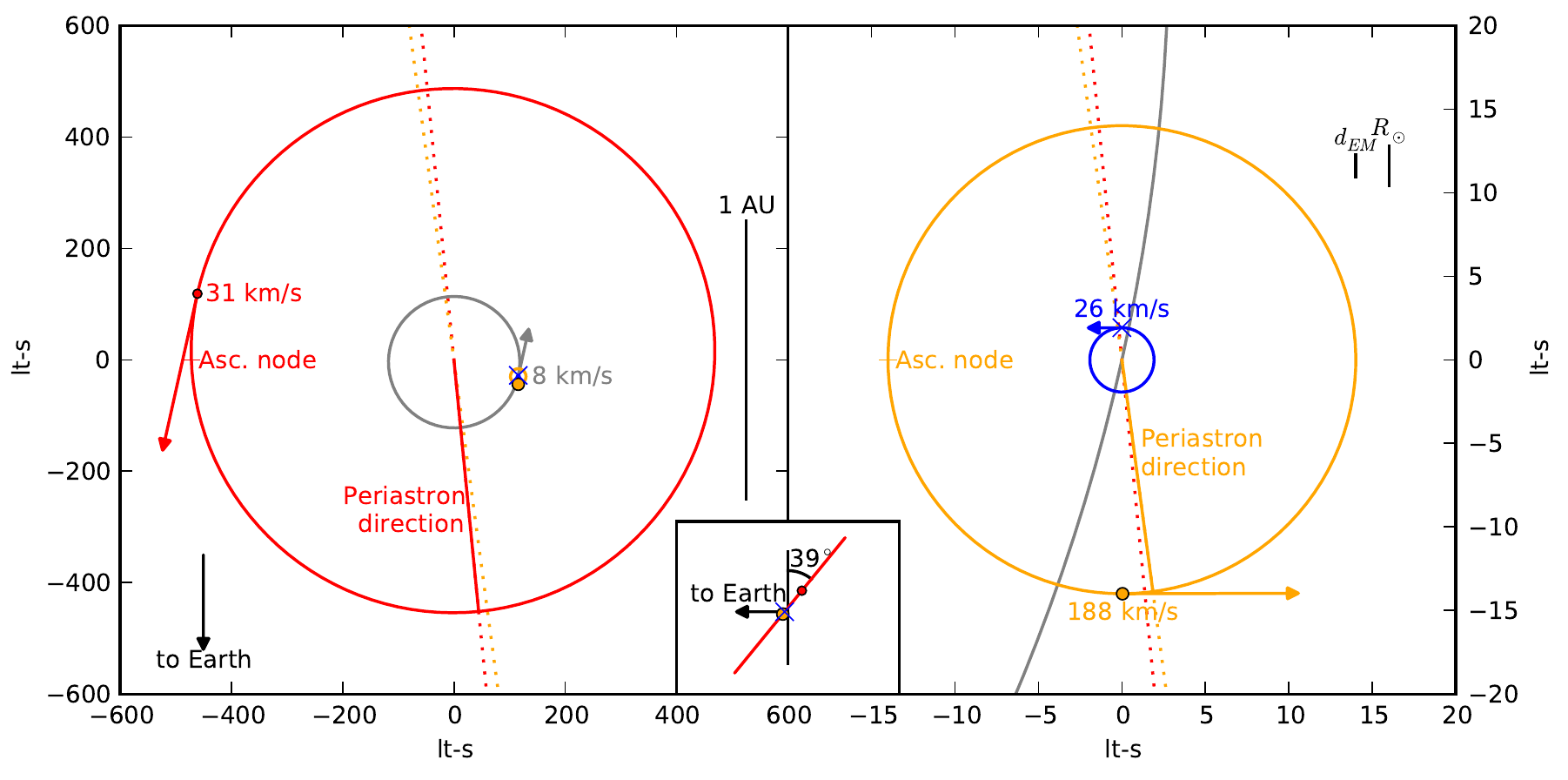}}
  \caption{Geometry of the PSR~\psr\ system at the reference epoch.
    The left panel shows the orbital shape and velocity of the outer
    white dwarf (red) and the orbital shape and velocity of the centre
    of mass of the inner binary (grey).  The right panel shows the
    orbital shapes and velocities of the inner white dwarf (orange)
    and pulsar (blue).  Dotted red and orange lines indicate the
    directions of periastron for the inner and outer white dwarf
    orbits, respectively.  The white dwarf positions when the pulsar
    or inner orbit center of mass crosses the ascending nodes are
    indicated. Vertical lines show length scales in the system in
    Astronomical Units (AU; left), or the Earth-Moon distance and the
    Solar radius ($d_{\rm EM}$ and \rsun; right).  The inset plot at
    bottom centre shows the inclination of the basically coplanar
    orbits with respect to the Earth-pulsar
    direction.\label{fig:diagram}}
\end{figure}

To fold the pulsar signal we approximate the motion of \psr\ using a
pair of Keplerian orbits, with the centre of mass of the inner orbit
moving around in the outer orbit.  We determine pulse times of arrival
(TOAs) from the folded radio data using standard techniques (Methods)
and then correct them to the Solar System barycentre at infinite
frequency using a precise radio position obtained with the Very Long
Baseline Array (VLBA; Methods).  These TOAs vary significantly
compared to a simple pulsar spin-down model by the R\o mer and
Einstein delays\cite{bh86}.  The R\o mer delay is a simple geometric
effect due to the finite speed of light and therefore measures the
pulsar's orbital motion. Its amplitude is $a_I\sin i/c \sim 1.2$\,sec
for the inner orbit and $\sim$74.6\,sec for the outer orbit (see
Figures~\ref{fig:resids} and \ref{fig:diagram}).

The Einstein delay is the cumulative effect of time dilation, both
special-relativistic due to the transverse Doppler effect, and
general-relativistic via gravitational redshift due to the pulsar's
position in the total gravitational potential of the system. For \psr,
the gravitational redshift portion is covariant with fitting the
projected semimajor axis of the orbit just as the full Einstein delay
is for other pulsars with circular orbits. The transverse Doppler
effect is easily measurable, though, since it is proportional to
$v^2/c^2 = |\mathbf{v}_I + \mathbf{v}_O|^2/c^2 = (v_I^2 + v_O^2 + 2
\mathbf{v}_I \cdot \mathbf{v}_O)/c^2$, where $\mathbf{v}_I$ and
$\mathbf{v}_O$ are the 3-dimensional velocities in the inner and outer
orbit, respectively, and $v_I$ and $v_O$ the corresponding speeds. The
$v^2$ terms are covariant with orbital fitting as in the binary case,
but the cross term $\mathbf{v}_I \cdot \mathbf{v}_O/c^2$ contributes
delays of tens of $\mu$s on the timescale of the inner orbit.

The two-Keplerian-orbit approximation results in systematic errors of
up to several hundred $\mu$s over multiple timescales due to unmodeled
three-body interactions (see Figure~\ref{fig:resids}), but those
systematics carry a great deal of information about the system masses
and geometry. The planet pulsar B1257$+$12 showed similar
systematics\cite{wf92} and direct numerical integrations confirmed
their planetary nature and provided the masses and orbits of the
planets\cite{peale93,wol94}.  The interactions in \psr\ are many
orders of magnitude stronger, but they, along with the R\o mer and
Einstein delays, can be similarly modeled by direct integration.

We use Monte Carlo techniques (Methods) to find sets of parameter
values that minimize the difference between measured TOAs and
predicted TOAs from three-body integrations, and we determine their
expected values and error estimates directly from the parameter
posterior distributions.  We plot the results in
Figure~\ref{fig:resids} and list best-fit parameters and several
derived quantities in Table~\ref{tab:params}.  Component masses and
relative inclinations are determined at the 0.1$-$0.01\% level, one to
two orders of magnitude more precisely than from other MSP timing
experiments, in what is effectively a gravitational-theory-independent
way.  A detailed description of the three-body model and fitting
procedure us under way (A.M.A.~\emph{et~al.}, manuscript in
preparation).

Based on an early radio position, we identified an object with
unusually blue colours in the Sloan Digital Sky Survey (SDSS;
Figure~\ref{fig:image})\cite{aaa+09c}.  The optical plus archival
ultraviolet photometry, combined with new near- and mid-infrared
photometry, are consistent (Methods) with a single $\sim$15,000\,K WD,
which optical spectroscopy confirmed is the inner WD in the system
(D.L.K.~\emph{et~al.}, manuscript in preparation).  When combined with
the known WD mass from timing, WD models provide a radius allowing us
to infer a photometric distance to the system of 1,300$\pm$80\,pc.
The photometry and timing masses also exclude the possibility that the
outer companion is a main sequence star.

The pulsar in this system appears to be a typical radio MSP, but it is
unique in having two WD companions in hierarchical orbits.  While more
than 300 MSPs are known in the Galaxy and in globular clusters, \psr\
is the first MSP stellar triple system found.  As there are no
significant observational selection effects discriminating against the
discovery of pulsar triple (as opposed to binary) systems, this
implies that $\lesssim$1\% of the MSP population resides in stellar
triples and that $\lesssim$100 such systems exist in the Galaxy.

Predictions for the population of MSP stellar triples have suggested
most would have highly eccentric outer orbits due to dynamical
interactions between the stars during stellar evolution\cite{pz+11}.
Such models can also produce eccentric binaries like MSP
J1903$+$0327\cite{crl+08}, if the inner WD, which had previously
recycled the pulsar, was destroyed or ejected from the system
dynamically\cite{fbw11}.  In these situations, however, the
co-planarity and circularity of the orbits of \psr\ would be very
surprising.  Those orbital characteristics, and their highly
hierarchical nature ($P_{b,O}/P_{b,I}$$\sim$200), imply that the
current configuration is stable on long timescales\cite{mar13},
greatly increasing the odds of observing a triple system like \psr.
Secular changes to the various orbital parameters will occur in the
long term\cite{fkr00}, however, and the three-body integrations and
timing observations will predict and measure them.

The basic evolution of the system, which was almost certainly complex
and exotic, may have progressed as follows.  The most massive of the
progenitor stars evolved off the main-sequence and exploded in a
supernova, creating the neutron star.  At least two of the companions
to the original primary survived the explosion, probably in eccentric
orbits.  After of-order 10$^9$\,years, the outermost star, the next
most massive, evolved and transferred mass onto the inner binary,
comprising the neutron star and a lower mass main sequence star,
perhaps within a common envelope.  During this phase, the angular
momentum vectors of the inner and outer orbits were torqued into near
alignment\cite{lp97}.  After the outer star ejected its envelope to
become a WD and another of-order 10$^9$\,years passed, the remaining
main-sequence star evolved and recycled the neutron star via the
standard scenario\cite{ts99}.  During this phase, the inner orbit
became highly circular but only a small amount of mass ($<$0.2\,\msun,
in total) was transferred to the neutron star: enough to drastically
increase its rotation rate, but not enough to make it particularly
massive\cite{opns12,lat12}.  Since then, secular effects due to
three-body interactions have aligned the apsides of the two
orbits\cite{fab11}, although our three-body integrations display
librations around apsidal alignment on both outer-orbital and secular
timescales.  This scenario nicely explains the co-planarity and
circularity of the orbits as well as the fact that both WD companions
fall on the predicted He-WD-mass vs.~orbital period
relation\cite{ts99}.

Perhaps the most interesting aspect of \psr\ is its potential to
provide extremely sensitive tests of the strong equivalence principle
(SEP)\cite{will06,fkw12}, a key implication of which is that the
orbital motions of bodies with strong self-gravity are the same as
those with weak self-gravity. In the case of \psr, a neutron star with
a gravitational binding energy $3GM/5Rc^2\sim0.1$ and a low-mass WD
with much smaller gravitational binding energy
($\sim$3$\times$10$^{-6}$) are both falling in the relatively strong
gravitational potential of the outer WD.  The five orders of magnitude
difference in the gravitational binding energies of the pulsar and WD,
as well as the large absolute value for the pulsar, provide a much
larger ``lever arm'' for testing the SEP than Solar System tests,
where the planets and moons have gravitational binding energies
between $10^{-11}$ and $10^{-9}$. Previous strong-field SEP tests used
MSP$-$WD systems and the Galactic field as the external perturbing
field\cite{sfl05,gsf11}. In the case of \psr, the perturbing field
(that of the outer WD) is 6$-$7 orders of magnitude larger, greatly
magnifying any possible SEP violation effects\cite{will06,fkw12}.
Since most metric theories of gravity besides general relativity
predict violations of the SEP at some level, high-precision timing of
\psr\ should soon produce unique and extremely interesting new tests
of gravity\cite{will06}.

\section*{METHODS SUMMARY}

We have taken many hundreds of hours of radio timing observations with
the GBT, Arecibo, and WSRT over the last two years, with the best
observations having time-of-arrival uncertainties of 0.8\,$\mu$s in
10\,s of data.  These TOAs are fit using a high-precision numerical
integrator, including Newtonian gravitational effects from the
three-body interactions as well as special-relativistic transverse
Doppler corrections and general-relativistic Einstein and Shapiro
delays (the last two are not yet important for the fitting).
Uncertainties on the fitted parameters are derived using Markov chain
Monte Carlo (MCMC) techniques.

The optical/ultraviolet/infrared photometry of the inner object were
fit using an absorbed WD atmosphere, finding results very close to the
spectroscopic values.  Based on the spectroscopic gravity, the
photometric radius of the inner WD is 0.091$\pm$0.005\,$R_\odot$ which
leads to a photometric distance to the system.  We see no emission
from the outer object, and can reject all single or binary
main-sequence stars as being the outer companion. The data are
consistent with a 0.4\,$M_\odot$ WD.

The radio timing fits benefitted from a radio interferometric position
of \psr\ determined from a 3-hour observation with the VLBA.  The
absolute positional accuracy is estimated to be 1$-$2\,milli-arcsec.
A series of observations which has already begun will determine the
parallax distance to 1$-$2\% precision as well as the 237/$D_{\rm
  kpc}$\,$\mu$-arcsec reflex motion on the sky caused by the outer
orbit, where $D_{\rm kpc}$ is the distance to the system in kpc.
\vspace{18pt}




\begin{addendum}
\item We thank D.~Levitan and R.~Simcoe for providing optical and
  infrared observations, J.~Deneva for early Arecibo observations,
  P.~Bergeron for use of his white dwarf photometry models; K.~O'Neil
  and F.~Camilo for approving discretionary time observations on the
  GBT and Arecibo, respectively, J.~Heyl, E.~Algol, and P.~Freire for
  discussions, and G.~Kuper, J.~Sluman, Y.~Tang, G.~Jozsa, and
  R.~Smits for their help supporting the WSRT observations.  The GBT
  and VLBA are operated by the National Radio Astronomy Observatory, a
  facility of the National Science Foundation operated under
  cooperative agreement by Associated Universities, Inc. The Arecibo
  Observatory is operated by SRI International in alliance with Ana
  G.~M\'endez-Universidad Metropolitana and the Universities Space
  Research Association, under a cooperative agreement with the
  National Science Foundation.  The WSRT is operated by the
  Netherlands Institute for Radio Astronomy (ASTRON).  This paper made
  use of data from the WIYN Observatory at Kitt Peak National
  Observatory, National Optical Astronomy Observatory, which is
  operated by the Association of Universities for Research in
  Astronomy (AURA) under cooperative agreement with the National
  Science Foundation. This work is also based in part on observations
  made with the Spitzer Space Telescope, which is operated by the Jet
  Propulsion Laboratory, California Institute of Technology under a
  contract with NASA.  I.H.S., V.M.K., M.H.v.K., and A.B.~acknowledge
  support from NSERC.  A.M.A.~and J.W.T.H.~acknowledge support from a
  Vrije Competitie grant from NWO.  J.B., D.R.L, V.I.K., \&
  M.A.M.~were supported by a WV EPSCoR Research Challenge Grant.
  V.M.K.~acknowledges support from CRAQ/FQRNT, CIFAR, Canada Research
  Chairs Program, and the Lorne Trottier Chair.
\item[Contributions] S.M.R., M.A.M., and D.R.L. were co-PIs of the GBT
  survey which found the pulsar, and all other other authors except
  D.L.K., M.H.v.K., A.T.D., S.C., and A.S.-R. were members of the
  survey team who observed and processed data.  J.B.~found the pulsar
  in the search candidates.  S.M.R.~identified the source as a triple,
  wrote follow-up proposals, observed with the GBT, phase-connected
  the timing solution, and wrote the manuscript.  I.H.S.~and J.W.T.H.~
  performed timing observations, wrote follow-up proposals, and
  substantially contributed to the initial timing solution.
  A.M.A.~developed the successful timing model and performed the
  numerical integrations and MCMC analyses.  D.L.K.~identified the
  optical counterpart and then with M.H.v.K.~and A.S.-R.~performed
  optical/IR observations and the multi-wavelength analysis.
  M.H.v.K.~and D.L.K.~both helped develop parts of the timing model.
  A.T.D.~and S.C.~performed the VLBA analysis.  All authors
  contributed to interpretation of the data and the results and the
  final version of the manuscript.
\item[Competing Interests] The authors declare that they have no
competing financial interests.
\item[Correspondence] Correspondence and requests for materials should
be addressed to S.M.R.~(email: sransom@nrao.edu).
\end{addendum}



\begin{table}
  \spacing{1}
  \begin{center}
    \footnotesize
    \begin{tabular}{lcc}
Parameter & Symbol & Value \\
\hline
\multicolumn{3}{c}{Fixed values}\\
\hline
Right ascension & RA & $03^h 37^m 43^s.82589(13)$\\
Declination & Dec & $17\degree 15' 14'' .828(2)$\\
Dispersion measure & DM & $21.3162(3)$\,pc\,cm$^{-3}$\\
Solar system ephemeris &  & DE405\\
Reference epoch &  & MJD $55920.0$\\
Observation span &  & MJD $55930.9-56436.5$\\
Number of TOAs &  & $26280$\\
Weighted root-mean-squared residual &  & $1.34\,\mu$s\\
\hline
\multicolumn{3}{c}{Fitted parameters}\\
\hline
\multicolumn{3}{c}{Spin-down parameters}\\
Pulsar spin frequency & $f$ & $365.953363096(11)$ Hz\\
Spin frequency derivative & $\dot f$ & $-2.3658(12)\times 10^{ -15 }$ Hz s$^{-1}$\\
\multicolumn{3}{c}{Inner Keplerian parameters for pulsar orbit}\\
Semimajor axis projected along line of sight & $(a \sin i)_I$ & $1.21752844(4)$ lt-s\\
Orbital period & $P_{b,I}$ & $1.629401788(5)$ d\\
Eccentricity parameter $(e\sin \omega)_I$ & $\epsilon_{1,I}$ & $6.8567(2)\times 10^{ -4 }$ \\
Eccentricity parameter $(e\cos \omega)_I$ & $\epsilon_{2,I}$ & $-9.171(2)\times 10^{ -5 }$ \\
Time of ascending node & $t_{\text{asc},I}$ & MJD $55920.407717436(17)$\\
\multicolumn{3}{c}{Outer Keplerian parameters for centre of mass of inner binary}\\
Semimajor axis projected along line of sight & $(a \sin i)_O$ & $74.6727101(8)$ lt-s\\
Orbital period & $P_{b,O}$ & $327.257541(7)$ d\\
Eccentricity parameter $(e\sin \omega)_O$ & $\epsilon_{1,O}$ & $3.5186279(3)\times 10^{ -2 }$ \\
Eccentricity parameter $(e\cos \omega)_O$ & $\epsilon_{2,O}$ & $-3.462131(11)\times 10^{ -3 }$ \\
Time of ascending node & $t_{\text{asc},O}$ & MJD $56233.935815(7)$\\
\multicolumn{3}{c}{Interaction parameters}\\
Semimajor axis projected in plane of sky & $(a \cos i)_I$ & $1.4900(5)$ lt-s\\
Semimajor axis projected in plane of sky & $(a \cos i)_O$ & $91.42(4)$ lt-s\\
Inner companion mass over pulsar mass & $q_I = m_{cI}/m_p$ & $0.13737(4)$ \\
Difference in longs. of asc. nodes & $\delta_\Omega$ & $2.7(6)\times 10^{ -3 }$ $\degree$\\
\hline
\multicolumn{3}{c}{Inferred or derived values}\\
\hline
\multicolumn{3}{c}{Pulsar properties}\\
Pulsar period & $P$ & $2.73258863244(9)$ ms\\
Pulsar period derivative & $\dot P$ & $1.7666(9)\times 10^{ -20 }$ \\
Inferred surface dipole magnetic field & $B$ & $2.2\times 10^8$ G\\
Spin-down power & $\dot E$ & $3.4\times 10^{34}$ erg s$^{-1}$\\
Characteristic age & $\tau$ & $2.5\times 10^9$ y\\
\multicolumn{3}{c}{Orbital geometry}\\
Pulsar semimajor axis (inner) & $a_I$ & $1.9242(4)$ lt-s\\
Eccentricity (inner) & $e_I$ & $6.9178(2)\times 10^{ -4 }$ \\
Longitude of periastron (inner) & $\omega_I$ & $97.6182(19)$ $\degree$\\
Pulsar semimajor axis (outer) & $a_O$ & $118.04(3)$ lt-s\\
Eccentricity (outer) & $e_O$ & $3.53561955(17)\times 10^{ -2 }$ \\
Longitude of periastron (outer) & $\omega_O$ & $95.619493(19)$ $\degree$\\
Inclination of invariant plane & $i$ & $39.243(11)$ $\degree$\\
Inclination of inner orbit & $i_I$ & $39.254(10)$ $\degree$\\
Angle between orbital planes & $\delta_i$ & $1.20(17)\times 10^{ -2 }$ $\degree$\\
Angle between eccentricity vectors & $\delta_\omega \sim \omega_O-\omega_I$ & $-1.9987(19)$ $\degree$\\
\multicolumn{3}{c}{Masses}\\
Pulsar mass & $m_p$ & $1.4378(13)$ $M_\Sun$\\
Inner companion mass & $m_{cI}$ & $0.19751(15)$ $M_\Sun$\\
Outer companion mass & $m_{cO}$ & $0.4101(3)$ $M_\Sun$\\
\hline
\end{tabular}

  \end{center}
\end{table}


\begin{table}
  \caption {\footnotesize 
    System parameters for PSR~\psr. Values in parentheses represent
    1-$\sigma$ errors in the last decimal place(s), as determined by MCMC
    fitting (see Methods). The top section contains parameters supplied as 
    input to our
    timing fit; the position was obtained from an observation with the
    VLBA, and the DM was measured from high-signal-to-noise
    Arecibo observations.  The middle section contains the parameters used to
    describe the state of the system at the reference epoch -- the initial
    conditions for the differential equation integrator. Along
    with the pulsar spin-down parameters, these parameters include the
    conventional Keplerian elements measurable in binary pulsars for each
    of the orbits, plus four parameters measurable only due to three-body
    interactions. These fourteen parameters can completely describe any 
    configuration of three masses, positions, and velocities for which the 
    center of mass remains fixed at the origin, provided that the longitude 
    of the inner ascending node is zero. Although some fitting parameters 
    are highly covariant, we computed all parameters and their errors 
    based on the posterior distributions, taking into account these
    covariances. We use the
    standard formulae for computing $B$, $\dot E$, and $\tau$, which assume 
    a pulsar mass of 1.4\,\msun\ and a moment of inertia of 
    $10^{45}$\,g\,cm$^2$. We have not corrected these values for proper 
    motion or Galactic acceleration.  The Laplace-Lagrange parameters 
    $\epsilon_1$ and $\epsilon_2$ parameterize eccentric orbits in a way 
    that avoids a coordinate singularity at zero eccentricity. The pair
    $(\epsilon_2,\epsilon_1)$ forms a vector in the plane of the orbit
    called the eccentricity vector.  For a single orbit, the ascending
    node is the place where the pulsar passes through the plane of the sky
    moving away from us; the longitude of the ascending node specifies the
    orientation of the orbit on the sky. This is not measurable with the
    data we have, but the difference between the longitudes of the
    ascending nodes of the two orbits is measurable through orbital
    interactions.  The invariant plane is the plane perpendicular to the
    total (orbital) angular momentum of the triple system.
    \label{tab:params}}
\end{table}

\clearpage

\clearpage

\begin{figure}
  \centerline{\epsfig{angle=0.0,width=\hsize,
      file=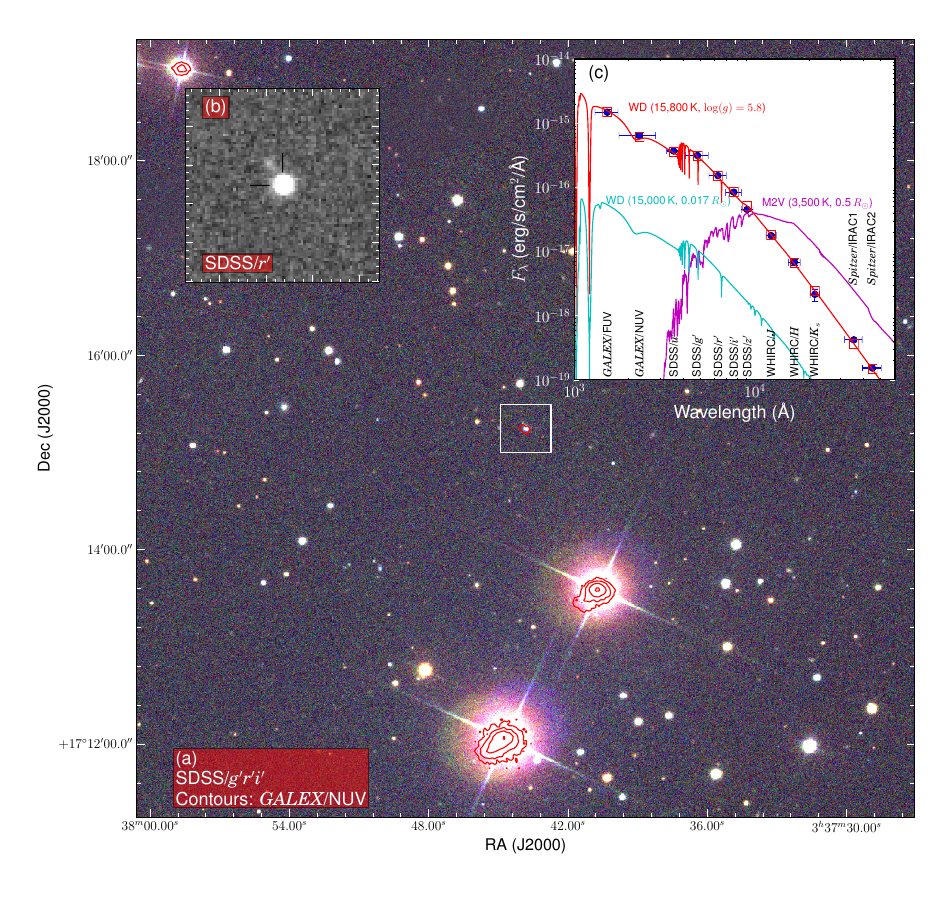}}
  \caption{Optical, infrared and ultraviolet data on PSR~\psr. The main
    panel (a) is a 3-colour optical image around \psr, from the Sloan
    Digital Sky Survey (Methods).  The contours are Galaxy Evolution
    Explorer (\textit{GALEX}) near-ultraviolet (NUV) data. The box at
    the centre is the area of the inset in the upper-left (b), which
    shows a $30^{\prime \prime}\times 30^{\prime\prime}$ region of the
    SDSS $r^\prime$ filter (SDSS/$r^\prime$) image along with the VLBA
    position indicated by the tick marks.  The inset in the
    upper-right (c) shows the spectral energy distribution.  The data
    from \textit{GALEX}, SDSS, the WIYN High Resolution Infrared
    Camera (WHIRC) and the Spitzer Infrared Array Camera (IRAC) are
    the blue circles, as labelled.  The error bars represent 1$\sigma$ 
    uncertainties in the $y$ direction (flux density) and the widths of 
    the photometric filters in the $x$ direction (wavelength).  
    The red curve is a model atmosphere
    consistent with our spectroscopic determination for the inner
    white dwarf (WD) companion.  The red boxes are the model atmosphere
    integrated through the appropriate filter passbands.  For
    comparison, we also plot two possible models for the outer
    companion: a M2V star\cite{kurucz93} (magenta) and a 0.4\,\msun\
    white dwarf (cyan).  All models have been reddened with an extinction of
    $A_V=0.44\,$mag. The photometry is consistent with the light from
    the hot inner white dwarf (with a surface gravity $\log(g)$=5.8) 
    and a smaller and more massive outer
    white dwarf, but not a low-mass main-sequence star.
    \label{fig:image}}
\end{figure}

\begin{methods}
  \subsection{Radio timing observations.}

  Continuing timing observations over most of the past two years have
  been undertaken with Arecibo every few weeks, the GBT every week to
  10 days, and the WSRT nearly daily, providing complementary timing
  precision and cadence.  Without a complete timing solution, nor even
  a rough estimate of outer orbit parameters for most of the duration,
  the rapid cadence was essential to maintain unambiguous count of the
  pulsar's rotations.  Similar centre frequencies (1,440\,MHz /
  1,500\,MHz), effective bandwidths ($\sim$600\,MHz / $\sim$700\,MHz)
  and observing systems (PUPPI / GUPPI)\cite{guppi08} are used at
  Arecibo and the GBT, respectively, while at the WSRT, the PUMAII
  instrument\cite{pumaii08} is used with $\sim$160\,MHz of bandwidth
  centred at 1,380\,MHz. In all cases the data are coherently
  de-dispersed\cite{hr75} at the pulsar's dispersion measure
  (21.3162(3)\,pc\,cm$^{-3}$) and folded modulo estimates of the
  predicted apparent pulsar spin period, initially determined from a
  polynomial expansion of inner orbital parameters re-fit on a weekly
  to monthly basis.

  Pulse times of arrival (TOAs), determined by cross-correlating high
  signal-to-noise template profiles with folded pulse profiles using
  standard practices\cite{taylor92}, are measured every 10 seconds to
  10 minutes depending on the telescope.  We use the precise VLBA
  position (see below) and {\tt TEMPO2}\cite{tempo2} to convert the
  arrival times at the observatory locations to the Solar System
  barycentre using the JPL DE405 planetary system ephemeris.

  \subsection{Timing fitting procedures and the three-body model.}

  High-precision three-body integrations determine the stellar masses
  and a nearly complete orbital geometry using only well-tested
  Newtonian gravity and special relativity.  For each set of trial
  parameters, we compute the pulsar and companion masses, positions,
  and velocities, and then use Newtonian gravity to compute their
  accelerations.  We evolve the system forward using a Bulirsch-Stoer
  differential equation solver\cite{bs66} (using 80-bit floating-point
  precision with {\tt ODEINT} in the {\tt Boost} library), obtaining
  position accuracy (limited by roundoff and truncation errors) on the
  order of a meter. We then compute the R\o mer and Einstein delays and
  use a spin-down model of the pulsar to produce a set of predicted
  TOAs, which we compare to the observed TOAs using a weighted sum of
  squared residuals.

  Fits to the measured TOAs without obvious systematic residuals are
  impossible without the inclusion of the three-body interactions as
  well as the special-relativistic transverse Doppler effect. General
  relativity is, in general, unimportant in the fitting of the system,
  but we calculate the full Einstein and Shapiro delays\cite{bh86}
  based on the determined system masses and geometry and incorporate
  them into the resulting best-fit parameters.  Ignoring these effects
  would lead to a distortion of orbital parameters, particularly the
  projected semimajor axes, as the delays would be absorbed into the
  fit.  The magnitude of the Shapiro delay is $\sim$2.9\,$\mu$s and
  $\sim$5.8\,$\mu$s peak-to-peak over the inner and outer orbits,
  respectively.

  The parameter space is explored using Markov chain Monte Carlo
  techniques\cite{emcee13}, and the parameter values given in Table~1
  are the Bayesian posterior expected values. We also use the
  posterior distribution to compute the standard deviations, quoted as
  1-sigma uncertainties.  This process marginalizes over covariances
  between parameters and derived values.

  \subsection{Ultraviolet, optical, and infrared observations.}

  After we identified \psr\ in the SDSS Data Release 7\cite{aaa+09c},
  we identified the same object as an ultraviolet source in the
  \textit{GALEX} All-sky Imaging Survey\cite{2007ApJS..173..682M},
  confirming the blue colours (Figure~\ref{fig:image}). We obtained
  further near-infrared photometry with the WHIRC imager\cite{whirc2}
  on the Wisconsin Indiana Yale NOAO (WIYN) 3.5-m telescope and
  mid-infrared photometry with the post-cryogenic Infrared Array
  Camera (IRAC) onboard the \textit{Spitzer Space
    Telescope}\cite{fha+04}.

  We fit the data using synthetic WD photometry\cite{tbg11} (extended
  to \emph{GALEX} bands by P.~Bergeron, personal communication),
  finding extinction $A_V=0.34\pm0.04$\,mag and effective temperature
  $T_{\rm eff}$=14,600$\pm$400\,K, with $\chi^2=7.7$ for 9 degrees of
  freedom (Figure~\ref{fig:image} inset (c)).  This is close to the
  effective temperature we determined via optical spectroscopy
  (D.L.K.~\emph{et~al.}, manuscript in preparation), $T_{\rm
    eff,spec}$=15,800$\pm$100\,K.  Radial velocity measurements from
  those spectroscopic observations confirm that the optical star is
  the inner WD in the system.  Given the spectroscopy-determined
  temperature and surface gravity of $\log g$=5.82$\pm$0.05, and a
  radius of 0.091$\pm$0.005\,\rsun\ based on the WD mass from pulsar
  timing (0.197\,\msun), the synthetic photometry provides a
  photometric distance to the system of 1,300$\pm$80\,pc.  That
  distance is somewhat larger than the $\sim$750\,pc implied by the
  measured DM toward the pulsar and the NE2001 Galactic free electron
  density model\cite{ne2001}, although the latter likely has a large
  error range.

  The photometry we measure is fully consistent with expectations for
  the inner companion only, as seen in Figure~\ref{fig:image}.  No
  additional emission is needed over the 1,000$-$50,000\,\AA\ range,
  although only where we actually have spectra can we be certain that
  no other emission is present.  Given its known mass of 0.4\,\msun,
  if the outer companion were a main-sequence star we would expect a
  spectral type of roughly M2V\cite{allen}, implying an effective
  temperature near 3,500\,K and a radius of 0.5\,\rsun.
  Figure~\ref{fig:image} shows such a stellar model\cite{kurucz93},
  which exceeds the near-IR and mid-IR data-points by a factor of
  $>$5, ruling out a main-sequence star as the outer companion.  Two
  main-sequence 0.2\,\msun\ stars would also cause an excess in the
  near- and mid-IR by a factor of $\sim$2.  Instead we find that a
  0.4\,\msun\ WD with effective temperature $<$20,000\,K (using
  synthetic photometry\cite{tbg11} again) is almost certainly the
  outer companion: such an object with $\log(g)$=7.5 and radius
  0.018\,\rsun\ leads to a change in the $\chi^2$ of 1 compared to the
  fit for only a single photosphere.  An example of such an outer
  companion is also shown in Figure~\ref{fig:image}.

  \subsection{Very Long Baseline Array Observations.}

  The position of \psr\ used in the timing analysis was determined
  from a single 3-hour observation with the Very Long Baseline Array
  (VLBA) on 13 February 2013, the first in a series of astrometric
  observations which will ultimately provide a $\sim$1$-$2\% parallax
  distance and transverse velocity for the system.  Eight
  dual-polarization, 32\,MHz wide subbands were sampled from within
  the range 1392$-$1712\,MHz, avoiding strong sources of radio
  frequency interference (RFI).  The bright source J0344$+$1559 was
  used as a primary phase reference source, and a phase referencing
  cycle time of 4.5\,minutes (total cycle) was employed.

  The multi-field correlation capability of the DiFX software
  correlator used at the VLBA\cite{deller11a} made it possible to
  inspect all catalogued sources from the NRAO VLA Sky Survey
  (NVSS)\cite{condon98a} which fell within the VLBA field of view; of
  the 44 such sources, 4 were detected by the VLBA and
  J033630.1$+$172316 was found to be a suitable secondary calibrator,
  with a peak flux density of 4\,mJy/beam.  The use of an ``in-beam''
  secondary calibrator reduces the spatial and temporal interpolation
  of the calibration solutions and improves the (relative) astrometric
  precision substantially\cite{chatterjee09a}.  \psr\ was detected
  with a signal-to-noise ratio of 30, providing a formal astrometric
  precision of around 0.1\,milli-arcsec.

  The absolute positional accuracy of \psr\ in the International
  Celestial Reference Frame is currently limited by the registration
  of the position of J033630.1$+$172316 relative to J0344$+$1559;
  given the angular separation of 2.3$\degree$ and the single
  observation, this is estimated at 1$-$2\,milli-arcsec.  This
  uncertainty in the absolute position will be reduced by additional
  VLBA observations.  In addition to improving the absolute position
  and solving for parallax and proper motion, a full VLBA astrometric
  model will incorporate the 237/$D_{\rm kpc}$\,$\mu$-arcsec reflex
  motion on the sky caused by the outer orbit, where $D_{\rm kpc}$ is
  the distance to the system in kpc.

  \subsection{Telescope Acknowledgements.}

\end{methods}

\end{document}